\documentclass[preprint2]{aastex}
\usepackage{amssymb}
\usepackage{epsfig}
\usepackage{natbib}
\usepackage{graphicx}

\begin{document}
\title{Influence of the nonlinearity parameter on the solar-wind sub-ion magnetic energy spectrum:
FLR-Landau fluid simulations}

\shorttitle{Solar wind energy spectrum at sub-ion scales}

\author{P.L. Sulem, T. Passot, D. Laveder and D. Borgogno}

\affil{Laboratoire Lagrange,\\
Universit\'e C\^ote d'Azur, CNRS,
Observatoire de la C\^ote d'Azur,}
\affil{CS 34229, 06304 Nice Cedex 4, France}
\email{sulem@oca.eu; passot@oca.eu; laveder@oca.eu; Dario.Borgogno@oca.eu}


\begin{abstract}

The cascade of kinetic Alfv\'en waves (KAWs) at the sub-ion scales 
in the solar wind is numerically simulated using a fluid approach
that retains ion and electron Landau damping, together with ion finite 
Larmor radius corrections. Assuming initially equal and isotropic ion and 
electron temperatures, and an ion beta equal to unity, 
different simulations are performed by varying the propagation direction and
the amplitude of KAWs that are randomly driven at a transverse scale of about one fifth
of the proton gyroradius in order to maintain a prescribed level of turbulent fluctuations. 
The resulting turbulent  regimes are characterized 
by the nonlinearity parameter, defined as the 
ratio of the characteristic times of Alfv\'en wave
propagation and of the transverse nonlinear dynamics.
The corresponding transverse magnetic energy spectra display  power 
laws  with exponents spanning  a range of values 
consistent with spacecraft observations.
The meandering of the magnetic field lines together with the ion temperature 
homogenization along these lines are 
shown to be related to the strength of the turbulence, measured by 
the nonlinearity parameter. 
The results are interpreted in terms of a recently proposed phenomenological
model where the homogenization process along field lines induced by Landau damping
plays a central role.

\end{abstract}

\keywords{ magnetic fields ---plasmas --- solar wind --- turbulence --- waves }



\section{Introduction}

An important issue when studying turbulence in the solar wind concerns  the 
energy spectrum $E(k_\perp)$ of the transverse magnetic fluctuations at sub-ion scales.
Spacecraft observations reveal a power law 
in a range of  scales  extending roughly  to the electron
gyroscale  $\rho_e$  \citep{ALM08,ACS13,BC13},  but the question arises of the 
universality of the spectral exponent \citep{Sah09,Sah10,Sah11,Alexandrova12,Chen13}. 
Observational values reported in Fig. 5 of  \citet{Sah13} are distributed within an interval $[-3.1,-2.5]$
with a peak of probability near $-2.8$, a range whose extension significantly exceeds 
the dispersion usually reported for the MHD-range spectral exponent,  evaluated for example
as $-1.63\pm 0.14$ in \citet{Smith06}. It turns out that, in some instances, the
MHD and sub-ion ranges
are separated by a short ``transition range'' located near the proton gyroscale 
$\rho_i$ where the spectrum
displays a power-law  with an exponent close to $-4$. The origin of this
range is not fully understood, although the
effect of monochromatic waves and coherent structures such as Alfv\'en vortices has recently been suggested 
(Alexandrova \& Lion, private communication).

Clearly, the  power-law observed for the transverse magnetic energy spectrum at the sub-ion scales 
is  steeper than the $k_\perp^{-7/3}$ prediction obtained by assuming a purely inertial cascade and
adapting to the ion scales the critical
balance argument \citep{GS95,NS11} that leads to the $k_\perp^{-5/3}$ spectrum in the MHD range.
Although it cannot be completely excluded that the observed variations of the spectral index 
originate from difficulties in the use of the Taylor hypothesis 
(on which most of the analysis of observational data are based)  beyond the ion scales, 
or from  a possible influence of the solar wind velocity and of its angle with the ambient field  
(Sahraoui \& Huang, private communication),  we shall here  focus on physical 
processes amenable to numerical simulations in the framework of fully kinetic descriptions or 
suitable fluid models.

Gyrokinetic simulations presented in  \citet{Howes11b} display  a sub-ion magnetic energy spectrum exponent very
close to the $-2.8$ most probable value reported from solar wind observations.
A more recent simulation performed with a different code shows a $-3.1$ sub-ion range \citep{Told15}.  
Although retaining only the low-frequency dynamics, such simulations require huge computational resources, 
making a parametric study difficult. 

On the other hand, reduced fluid models,
assuming isothermal ions and electrons, lead to a $k_\perp^{-8/3}$ spectrum,
where the departure from the $-7/3$ exponent has been related to intermittency corrections
originating from coherent structures \citep{Boldyrev12,Boldyrev13}. A $-8/3$ exponent is also obtained
in numerical simulations of electron-MHD in the presence of a strong ambient field 
\citep{Meyrand13}. In addition, energy spectra steeper than $-7/3$,
associated with coherent structures and deformation of the particle distribution functions,
are reported from fully kinetic particle-in-cell (PIC) simulations  with  a reduced mass 
ratio \citep{Wan15}, and from hybrid Eulerian Vlasov-Maxwell models \citep{Servidio15}. 

It turns out that, as the cascade proceeds, kinetic Alfv\'en waves (KAWs) become compressible, making 
Landau damping relevant. A phenomenological model 
was presented in \citet{Howes08,Howes11} where, due to this effect,
the cascade is non conservative and the energy flux decays
exponentially as the cascade proceeds. This leads to an exponential multiplicative correction to the $k^{-7/3}$
spectrum which, under specific choices of parameters, can indeed look like a
steeper power law. The intrinsic non-universality of the observed power-law energy spectrum is however not 
really reproduced. This model was recently revisited in \citet{PS15} by noting that Landau damping 
does not only lead to dissipation of the wave energy along the cascade but also induces 
a homogenization process of fields such as the temperatures along the distorted magnetic field lines, 
which affects the characteristic transfer time and thus the energy cascade.
It in particular modifies the power-law exponent that appears to be 
sensitive to the nonlinearity  parameter defined as the ratio of the 
Alfv\'en wave period to the nonlinear stretching time. Non-universality due to Landau damping
was also reported in  three-dimensional PIC simulations of whistler turbulence \citep{Gary12}, 
and in fact turns out to be a generic property of systems for which the characteristic times
of dissipation and transfer display the same variation in terms of the wavenumber \citep{Bratanov13}.

The aim of the present paper is to address the question of the non-universality of the sub-ion 
magnetic energy spectrum for a proton-electron plasma with initially equal and isotropic 
equilibrium mean temperatures, using numerical simulations of the FLR-Landau fluid  
described in Sections 2 and 5 of  \citet{SP15} (and references therein). 
The paper is organized as follows. In Section 2, we briefly review the main features
of the FLR-Landau fluid model and  specify the numerical setting. Section 3 discusses diagnostic tools. 
The main results are presented in Section 4 which, on the basis of FLR-Landau fluid numerical simulations,
correlates the sub-ion exponent of the transverse magnetic spectrum to the nonlinearity parameter.
Section 5 analyzes the meandering of magnetic field lines in various turbulence regimes. Section 6 provides an
interpretation of the numerical observations, based on a recently developed phenomenological model
\citep{PS15}. An important ingredient of this latter approach  is the homogenization process 
along the magnetic field lines of quantities such as the temperatures, which, in the case of ions but not of 
electrons, is supposed to affect the energy transfer. Section 7 demonstrates
that this homogenization effect is indeed captured in the numerical simulations. Section 8
is the Conclusion.

\section{The FLR-Landau fluid}

\subsection{Main features of the model}

The FLR-Landau fluid model appears as a system of dynamical equations for the proton density and 
velocity, the magnetic
field, the ion and electron parallel and perpendicular gyrotropic pressures and heat fluxes. 
The electric field is given by a generalized Ohm's law which includes the Hall term and
the electron pressure gradient, but neglects electron inertia. The fluid hierarchy
is closed by expressing the gyrotropic fourth-rank moments 
in terms of the above quantities, in a way consistent with 
the low-frequency linear kinetic theory. It in particular 
retains Landau damping and finite Lamor radius (FLR) corrections
(originating from the non-gyrotropic parts of the fluid moments and evaluated algebraically
in terms of the retained quantities), leading to an accurate description of the 
dispersion and dissipation of KAWs. 

It is of interest at this point to give a flavor of the way Landau damping is retained in such a fluid model.
A main assumption is that, up to the distortion of the magnetic field lines that is taken into account,
Landau damping keeps the same form as in the linear regime, an hypothesis probably valid 
when the turbulence fluctuations are not too strong \citep{Kanekar15}. 
Although the present simulations  involve a closure of the hierarchy at the 
level of the fourth-rank moments in order to retain a nonlinear dynamics for the gyrotropic
heat fluxes and also to provide a better description of Landau damping in the nonlinear regime
(see \citet{Schekochihin15} for a discussion), the physical interpretation is easier in the simplified framework
where the closure is done at the level of the heat fluxes. The equations for ion and electron
gyrotropic pressures involve terms of the form
$\nabla \cdot (q {\widehat {\mathbf b}})$ where $q$ is the parallel or perpendicular heat flux 
for the ions or the electrons, and 
${\widehat {\mathbf b}}$ the unit vector parallel to the local magnetic field. 
In a magnetized weakly collisional plasma, each of the gyrotropic heat fluxes is modeled as 
$q \sim {\widehat {\mathbf b}}\cdot \nabla {\widetilde T}$ where ${\widetilde T}$ 
refers to  the corresponding temperature fluctuations \citep{Spitzer-Harm} 
Differently, in the collisionless regime, the heat flux  scales like $v_{th}{\widetilde T}$
\citep{Hollweg74}, where $v_{th}$ is the 
associated thermal velocity. Transition between these two regimes in the solar wind was recently studied both from 
observational data \citep{Bale13} and numerical simulations \citep{Landi14}. The low-frequency linear kinetic theory
reproduces this scaling, but the relation involves a  Hilbert transform
along the equilibrium magnetic field lines, making  the system dissipative as
a consequence of Landau damping (see e.g. \citet{SHD97}).
In the nonlinear regime,  as discussed in \citet{Passot14}, the distortion of the magnetic field lines is to
be retained, requiring an approximation of the Hilbert transform along
distorted lines. Substituting the resulting expressions for the heat fluxes into the pressure equations
results in a homogenization process along the distorted magnetic lines,  through the 
operator (having the dimension of an inverse time scale) $v_{th} {\widehat b}\cdot \nabla {\cal H}$
where ${\cal H}$ refers to the Hilbert transform along the magnetic 
field line  (or in practice an approximation of it). 
The parallel or perpendicular ion or electron thermal velocity $v_{th}$ can be viewed as the rms 
particle streaming velocity and  the operator $v_{th} {\mathbf b} \cdot \nabla$ phenomenologically associated 
with a homogenization frequency $\omega_H \sim v_{th} k_\|$, where $k_\|$ 
refers to an inverse correlation length along the magnetic field lines. One may thus expect that the streaming
process is more efficient for the electrons than for the ions, and also when the strength of the turbulence 
(as measured by the nonlinearity parameter) is weaker, because the parallel wavenumber $k_\|$ is 
relatively smaller in this case, a point addressed in more details in Section 7.

\subsection{Numerical setting}

When numerically integrated, the FLR-Landau fluid equations given in \citet{SP15} 
are written in  non-dimensional variables, using  
the Alfv\'en velocity $v_A$,  the ion gyrofrequency $\Omega_i$
and the ambient magnetic field $B_0$ as basic units.
Furthermore, weak linear hyper-viscosity and hyper-diffusivity terms 
are supplemented in the equations for the velocity and magnetic field 
components respectively. In the spectral space, these terms involve
the same Fourier symbol $\nu (k_z^8 +  k_\perp^8)$
where the coefficient $\nu$ is  chosen as small
as possible, while permitting the development of a small-scale fast-decaying range for the energy spectrum
of the corresponding fields. These dissipative terms are added not only for preventing the development of 
numerical noise, but also to mimic the effect of the Landau dissipation 
at scales that are too small to be retained in the simulations. We checked that the resulting numerical 
dissipation is sufficiently small not to affect the power-law ranges. Terms 
corresponding to the work of the nongyrotropic pressure force are not retained in the equations for the 
ion parallel and perpendicular pressures. In a stationary regime where mean temperatures remain almost
constant, these terms have a minor effect. Nevertheless, the resulting model is not exactly conservative,
which does not appear to be a serious concern in situations involving both driving and numerical dissipation. 

\begin{table*}
\begin{center}
\def~{\hphantom{0}}
\begin{tabular}{|c||c|c|c|c|c|}
\hline
     &              &                  &              &                  &              \\
 Run & $R_{0.2}^{80}$ & $R_{0.13}^{80}$    & $R_{0.08}^{80}$ & $R_{0.08}^{83.6}$  & $R_{0.08}^{86}$ \\
     &              &                  &               &                 &      \\
\hline
$L_\perp/L_\|$ & 0.18 &0.18 & 0.18 & 0.11& 0.07 \\
\hline
Propagation angle of injected KAWs  & $80^\circ$   & $80^\circ$  & $80^\circ$ & $83.6^\circ$ & $86^\circ$\\
\hline
$\delta B_{\perp 0}/B_0 = {\langle |b_\perp|^2 \rangle }^{1/2}$ & 0.2 &  0.13 & 0.08 & 0.08 & 0.08 \\
\hline
$A= (k_z^{(0)}/k_\perp^{(0)})(B_0/\delta B_{\perp 0})$ & 0.9& 1.38 & 2.25 & 1.38 & 0.88 \\
\hline 
Transverse magnetic spectrum exponent & -2.3 & -2.6 & -3.6 & -2.8 & -2.3 \\
\hline
\end{tabular}
\caption{Run parameters and spectral exponents. Here $k_\perp^{(0)}$ and $k_z^{(0)}$ are the transverse and parallel
wavenumbers of the injected KAWs.}
\end{center}
\end{table*}

The FLR-Landau fluid equations are integrated using a Fourier pseudo-spectral method 
in a three-dimensional periodic domain, with a partial dealiasing ensured 
by spectral truncation at 2/3 of the maximal wavenumber in each coordinate direction. 
Parallelism is implemented using the so-called pencil decomposition.
In order to focus on the quasi-transverse dynamics, the spatial extension of the domain 
is larger in the parallel direction than in the transverse ones. 
Time integration is performed with an explicit $3^{\rm rd}$-order Runge-Kutta scheme.
A strong stability constraint on the time step originates from the use of a realistic ion-electron mass ratio,
while long integration times (typically involving $10^6$ time steps) 
are needed to reach a stationary state and obtaining significant 
statistical quantities in this regime.  For these reasons, the resolutions used 
at this stage were limited to  $128^3$ or $256^2 \times 128$ grid points. 

In the present simulations where the smallest perpendicular
wavenumber is $k_\perp d_i\approx 0.18$ (here $d_i$ denotes the ion inertial length), it is
reasonable, when motivated by the solar wind physics, to 
look at the largest retained scales as produced by the MHD cascade, and thus to assume that they 
display a significant scale anisotropy. As solar wind turbulence dominantly 
involves Alfv\'enic fluctuations, we are led to include in the
simulations, a driving that  mimicks the injection of  KAWs. For this purpose, 
we supplemented in the velocity equation  a random forcing with components
\begin{equation}
F_i(t,\mathbf{x}) = \sum_{1<n<N} F_{i,n}^{0} \ \cos(\omega_{_{KAW}}({\mathbf k}_n) t - 
{\mathbf k}_n \cdot {\mathbf x} + \phi_{i,n}),
\end{equation}
for different wavevectors ${\mathbf k}_n$, where $F_{i,n}^{0}$ and
$\phi_{i,n}$ are the amplitude and phase of the $n^{th}$ mode
of the $i^{th}$ component of the external driver $F_i$. When taking
for $\omega_{_{KAW}}({\mathbf k})$ the KAW frequency at wavevector
${\mathbf k}$, the system will generate KAWs by resonance at this frequency. KAW frequencies
are  calculated from the linearized FLR-Landau fluid  system, using
a symbolic mathematics software.
The dispersion relation is tabulated and used as an input in the FLR-Landau
fluid code. 
Eight KAWs (four forward and four backward propagating waves with an orthogonal polarization) 
with wavevector components $(\pm 2\pi /L_\perp, \pm 2\pi /L_\perp)$ ,  
$(\pm 2\pi /L_\perp, \mp 2\pi /L_\perp)$, $(0,\pm  2\pi /L_\perp)$  
and $(\pm  2\pi /L_\perp,0)$ in the transverse spectral plane and  $\pm 2\pi/L_\|$ 
in the parallel direction (where $L_\perp$ and $L_\|$ refer to the dimensions of 
the computing box) are excited, those at the largest scales propagating 
in a direction making a prescribed angle with the  ambient magnetic field, 
chosen from 80 to 86 degrees  by varying $L_\|$.
The driving is turned on (resp. off) when the sum of the
kinetic and magnetic energies is below (resp. above) prescribed
thresholds, taken such that the rms magnetic field fluctuations
remain at a prescribed level (from 0.057 to 0.2).
An alternative procedure, discussed in \citet{TenBarge14} and
implemented in gyrokinetic simulations \citep{Howes11b}, consists
in driving counterpropagating Alfv\'en waves by means of 
a parallel body current, using an oscillating antenna which obeys a Langevin equation.

Simulations were performed assuming initially equal isotropic ion and electron
temperatures and $\beta=1$ for each particle species. We use 
a computational domain of transverse size $L_{\perp}= 34.6\, d_i$,
and different longitudinal extensions $L_\|$, leading to different
propagation angles $\theta= {\rm atan} (L_\|/L_\perp)$ for the driven KAWs. 
The other parameter we vary is the rms  amplitude of the transverse 
(normalized) magnetic fluctuations $a=\delta B_{\perp0}/B_0$.
These quantities enters the definition of $A = k_z^{(0)}/k_\perp^{(0)} (B_0/\delta B_{\perp 0})
=(L_\perp / L_\|) (B_0/\delta B_{\perp 0})$, a ratio of two quantities that are
taken asymptotically small but of the same order in the gyrokinetic theory.
Table 1 summarizes the parameters of the various simulations, referred to as 
$R_a^\theta$. 

Due to the modest resolution of the present simulations, the retained wavenumbers 
in the transverse direction do not exceed a few  $\rho_i^{-1}$. 
Nevertheless, as these simulations do no include a transition range
which, as recalled in the Introduction, is not always observed in solar wind data, the 
power-laws for the transverse magnetic field obtained numerically can be viewed as characteristic 
of the sub-ion scales and extrapolated down to the electron gyroscale, thus permitting 
comparisons with the predictions of the phenomenological model introduced  in \citet{PS15}
and  discussed in Section 6.

\section{Diagnostic tools}

\subsection{Magnetic field lines} \label{magneticfieldlines}

In addition to the usual Eulerian quantities, it is useful to consider individual field lines at a time 
when turbulence has reached a steady state. We considered 32 such lines originating from points 
distributed over an array of $8 \times 4$ equispaced grid points 
in the range $10 \leq x \leq 24$, $10 \leq y \leq 24$, in the plane $z=0$. 
The magnetic field lines are computed by solving numerically the equation 
${d {\mathbf x}}/{ds}={\widehat {\mathbf b}}({\mathbf x})$, where $s$ is the curvilinear abscissa 
along a field line, using a standard variable-order, variable-step Adams method.
In order to make the computation 
of the field lines less time consuming, while keeping a reasonably good accuracy, we have adopted an approximate 
description of the magnetic field where only the 
Fourier modes of ${\mathbf B}$ with amplitude larger then $10^{-5}$ have been retained  \citep{BGPS08}. This
leads us to use of the order of $7 \times 10^4$ to $10^5$  Fourier modes for each scalar component in the strong 
turbulence regime and about $10^4$ for weak turbulence. 

\subsection{Parallel wavenumber} \label{paral-wave-sect}

An important issue, especially in the strong turbulence regime, is the distinction between the wavenumber $k_z$
along the uniform ambient magnetic field, and  the quantity $k_\|$ arising in the nonlinear regime. 
While the former can just be viewed 
as a Fourier variable, the latter is rather defined as the inverse correlation length 
of the amplitude of the magnetic field fluctuations along the 
magnetic field lines and, as such, turns out to be a function of the considered transverse scale. 
This leads to define the quantity $k_\|(k_\perp)$ for which a computation formula is given in \citet{CLV02}:
\begin{equation}
k_\|(k_\perp) \approx \left ( \frac{\sum_{k_\perp\le k'_\perp<k_\perp+\delta k_\perp} |\widehat{{\mathbf B}_L \cdot 
\nabla{\mathbf b}_l|^2_{\mathbf k'}}}{B_L^2\sum_{k_\perp\le k'_\perp<k_\perp+\delta k_\perp} |\widehat{{\mathbf b}_l}|^2_{\mathbf k'}}
\right )^{1/2}. \label{kparal}
\end{equation}
In Eq. (\ref{kparal}), ${\mathbf B}_L$ is the local mean field obtained by eliminating 
modes whose perpendicular wavenumber is greater than $k_\perp/2$, and ${\mathbf b}_l$ the 
fluctuating field obtained by eliminating
modes whose perpendicular wavenumber is less than $k_\perp/2$. In order to reduce statistical 
fluctuations on the evaluation of $k_\|$ when turbulence has reached a steady state, 
averaging is performed over 70 to 200 outputs, depending on the conditions.

\section{Nonlinearity parameter and spectral exponent} \label{spectr}

\begin{figure}
\begin{center}
\includegraphics[width=0.48\textwidth]{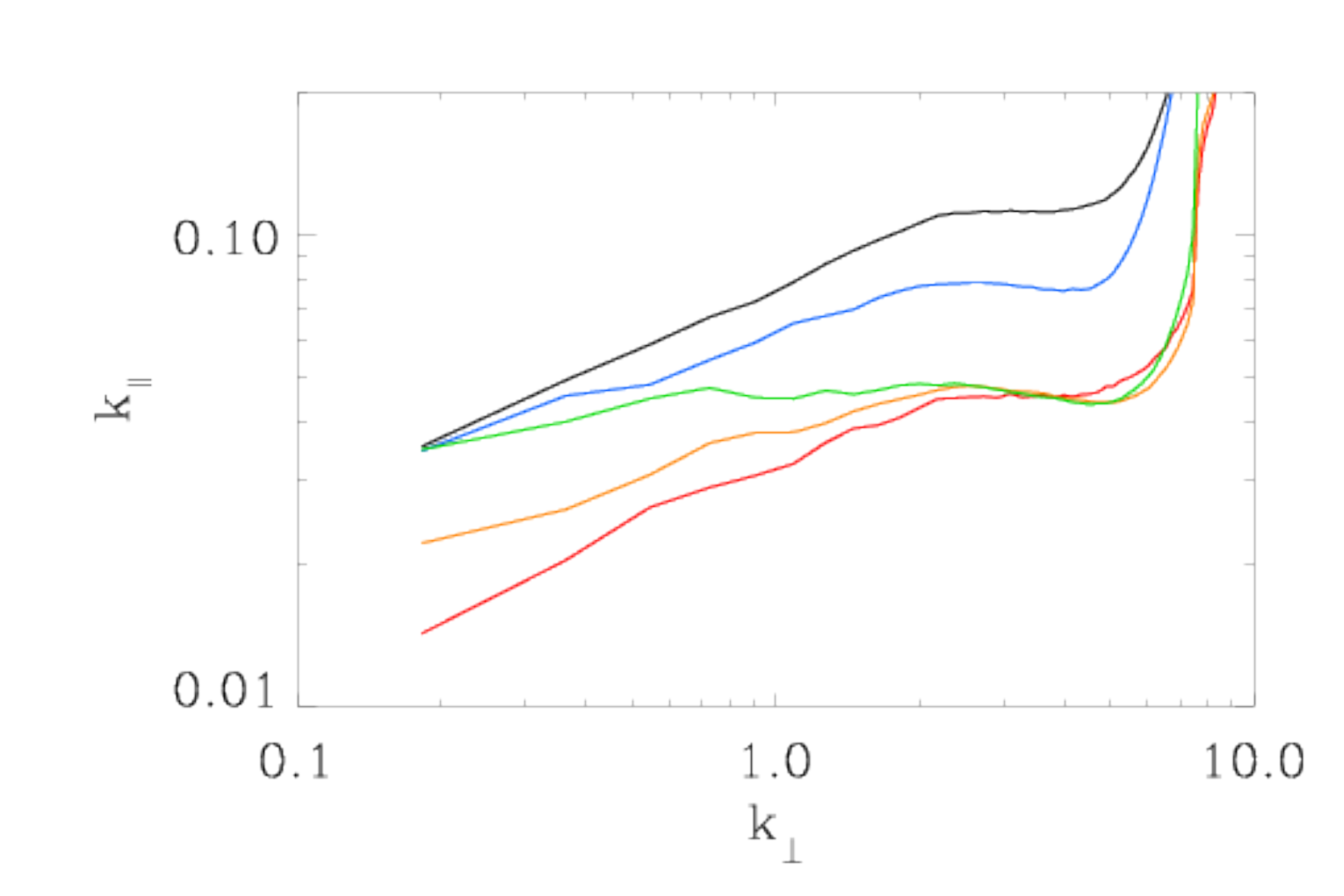}
\end{center}
\caption{Parallel wavenumber $k_\|(k_\perp)$ for runs $R_{0.2}^{80}$ (black), 
$R_{0.13}^{80}$ (blue), $R_{0.08}^{80}$ (green),  $R_{0.08}^{83.6}$ (orange) and
 $R_{0.08}^{86}$ (red).}
\label{kparal-fig}
\end{figure}

\begin{figure}
\begin{center}
\includegraphics[width=0.48\textwidth]{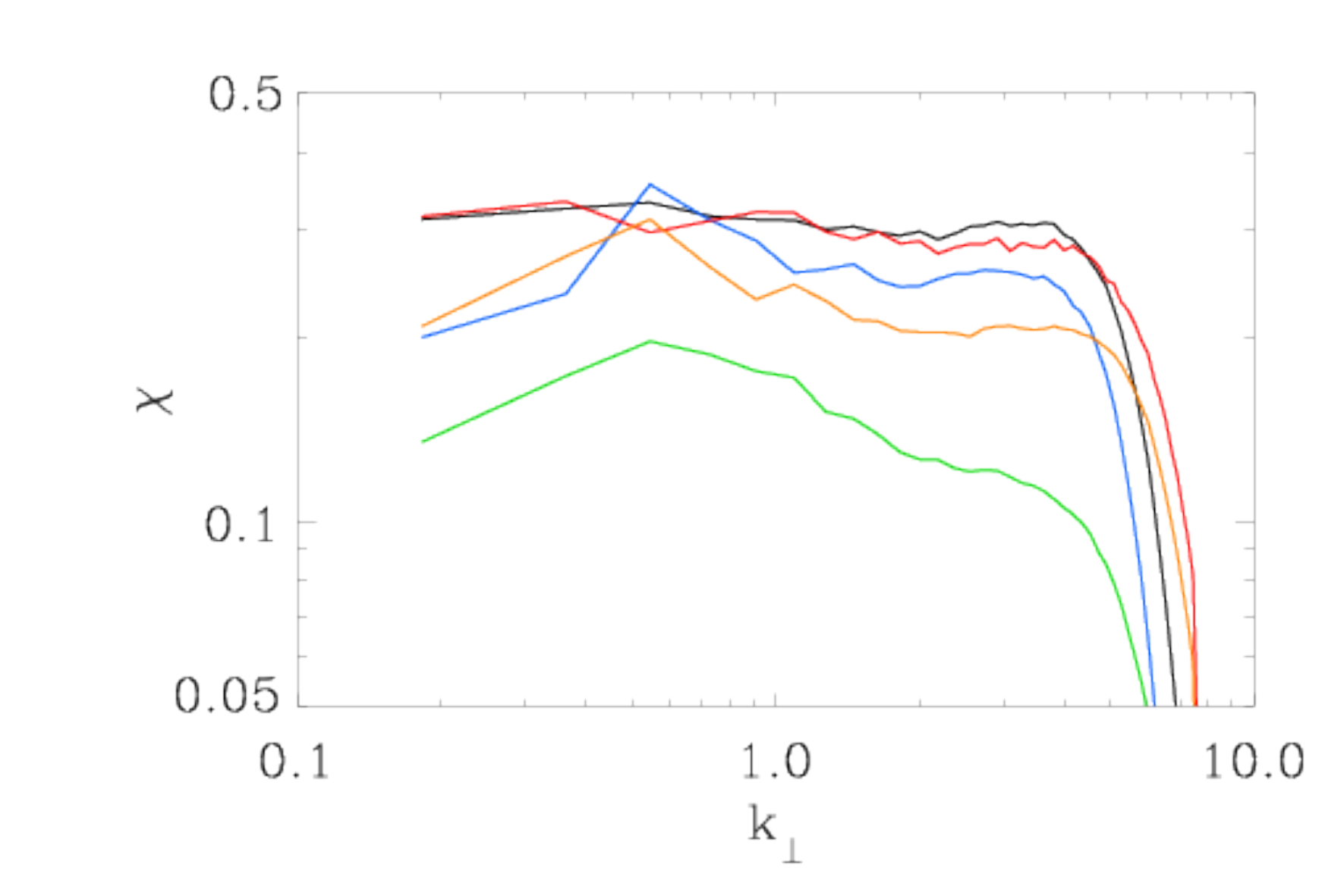}\\
\includegraphics[width=0.48\textwidth]{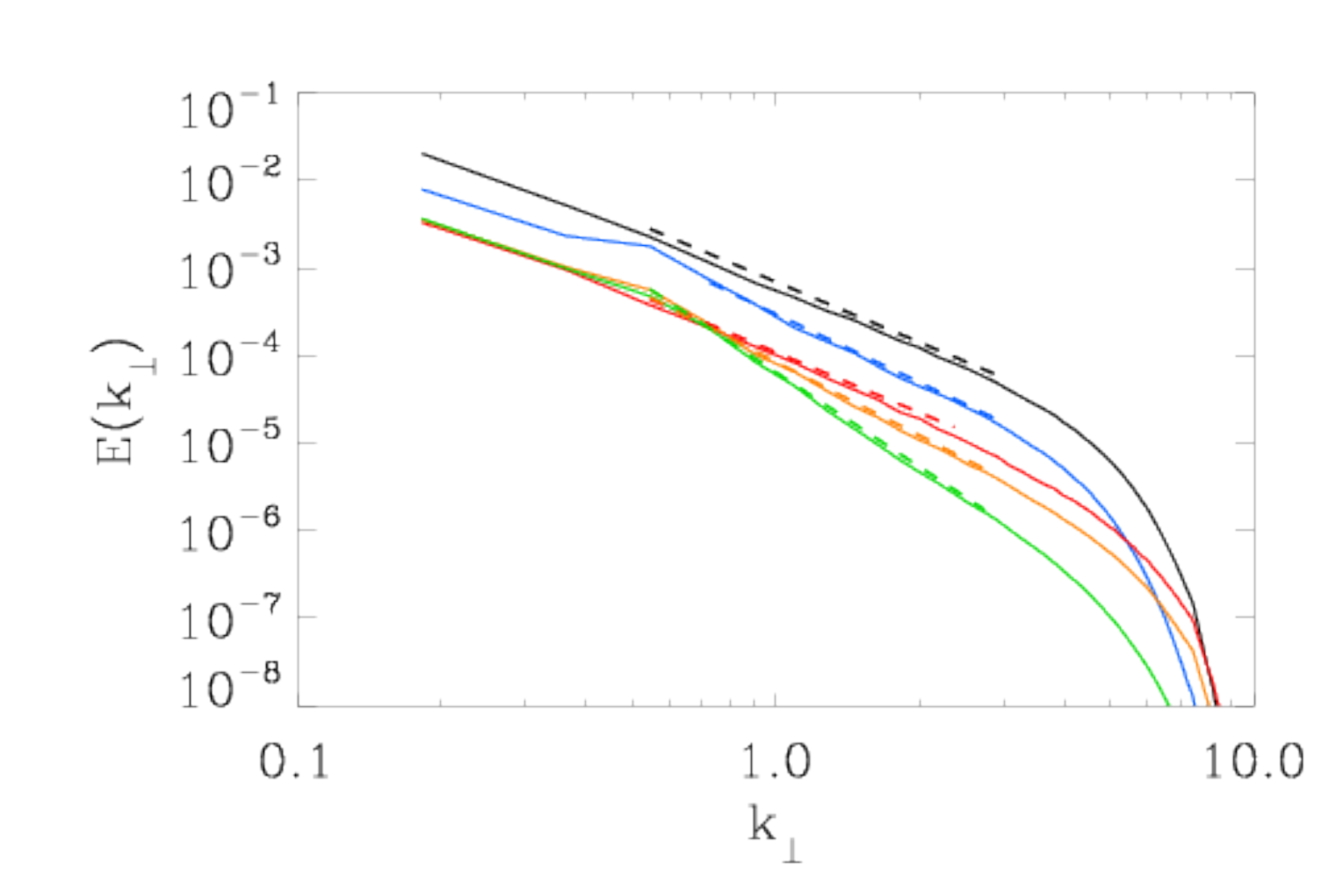}
\end{center}
\caption{Nonlinearity parameter $\chi$ (top) and transverse magnetic energy spectrum $E(k_\perp)$
(bottom) for the same runs as in Fig. \ref{kparal-fig} (same color code).}
\label{chi-spec}
\end{figure}

It is usually believed that strong turbulence in an anisotropic system in the presence of 
propagating waves is characterized by a ``critical balance'' condition where the characteristic
times of the transverse nonlinear dynamics  and of the wave propagation along the magnetic field lines
display the same scaling in terms of the transverse wavenumber $k_\perp$. This 
leads to define a ``nonlinearity parameter''
$\chi = \omega_{NL} / \omega_W$ as the ratio of the associated frequencies. 
In the present situation where the turbulent cascade is dominated by the
Alfv\'en wave dynamics, $\omega_W = {\overline \omega}(k_\perp\rho_i)k_\| v_A$. 
The function ${\overline \omega}$, which is provided by the linear kinetic theory, 
is equal to 1 at the MHD scales and  varies  like $k_\perp \rho_i$ when 
$k_\perp \rho_i \gg 1$. Furthermore, $\omega_{NL}$ corresponds to the rate of strain of the 
magnetic field by the electron velocity gradients. In the sub-ion range, electron velocity
(which is significantly larger than ion velocity)
scales approximately like the electric current.
When the energy spectrum $E(k_\perp)$ of the transverse magnetic fluctuations
is sufficiently shallow for local interactions to be dominant (which turns out to be the case
for $\beta$ of order unity), we estimate
$\omega_{NL} \sim {\overline \alpha} [k_\perp^3 E(k_\perp)]^{1/2}$, where 
${\overline \alpha}= {\overline \omega}$. Modeling of  non-local contributions is
presented in \citet{PS15}.

The longitudinal wavenumber $k_\|(k_\perp)$, computed as indicated in Section \ref{paral-wave-sect}, 
is displayed in Fig. \ref{kparal-fig} for the various simulations listed in Table 1. 
We note that  in the 
case of run $R_{0.08}^{80}$, $k_\|$ is almost constant (down to the scales
where hyperviscosity and hyperdiffusivity are significant and lead to a rapid increase), suggesting 
that this simulation corresponds to a weak turbulence regime. In contrast, 
for larger amplitude or larger propagation angle, $k_\|(k_\perp)$ 
grows as a power law and saturates only at smaller scales.

Figure \ref{chi-spec} (top) displays the nonlinearity parameter 
$\chi(k_\perp) = (k_\perp^3 E_k)^{1/2}/k_\|(k_\perp)v_A$ 
(assuming ${\overline \alpha}={\overline \omega}$), averaged over a time interval when the system has
reached a stationary state, for the various runs listed in Table 1. 
Two types of behavior can be distinguished.
In situations where the parameter $A$ is small, either because of a relatively large amplitude 
or a small ratio $L_\|/L_\perp$, the nonlinearity parameter $\chi(k_\perp)$  is essentially constant,
signature of a critically balanced turbulence.
Differently, in simulations for which $A$ is larger, $\chi(k_\perp)$ has a tendency to decrease.

Figure \ref{chi-spec} (bottom) displays the corresponding transverse 
energy spectra of the transverse magnetic fluctuations, 
averaged on the same time interval as the nonlinearity parameter. If at the smallest scales the spectra rapidly decay
due to the hyper-viscosity and hyperdiffusivity, one can clearly see
a power-law behavior in the close sub-ion range, in spite of the moderate extension of this range 
due to numerical constraints. An important observation is that the spectral exponent is not universal, but turns out to be 
correlated with the saturated value of the nonlinear parameter: as the latter is larger, the spectrum is shallower.

In connection with the above result, a few comments are in order. A first remark is that 
no transition range is present in the  simulations, suggesting that the same power-law spectrum extends from the ion to 
the electron gyroscale. Furthermore, a similar result is 
obtained by \citet{Gary12} in numerical simulations of whistler waves for which the power-law
spectrum at scales larger than the electron scales is reported to be shallower when, all other parameters being fixed,
the turbulence energy flux is increased. On the other hand, apparently different conclusions have been drawn from 
solar wind observational data. Recently, \citet{Bruno14} investigated the behavior of the spectral slope of 
interplanetary magnetic field fluctuations at proton scales from the WIND and MESSENGER spacecrafts at 1 AU and 0.56 AU,
respectively, moving from fast to slow wind regions. They confirm the variability of the spectral slope, but exponents
in the interval $-3.75$ to $-1.75$ are reported, with a tendency to approach $−5/3$ within the slowest wind, a value 
significantly in excess of theoretical and numerical predictions. They also
observed a tendency for the spectrum to be steeper where the speed is higher, and to be flatter within the subsequent
slower wind, following a gradual transition between these two states. They conclude that the
value of the spectral index depends on the power associated with the fluctuations within the inertial range,
the higher the power, the steeper the slope. Similar conclusions were reached by \citet{Smith06}. Comparisons with 
simulations where for example all the parameters are fixed except the amplitude of the turbulence fluctuations, is 
not straightforward. Indeed, as noted by \citet{Bruno14}, fluctuations within faster wind not only have
larger amplitudes but are also more Alfv\'enic,  i.e. display a larger cross-helicity. 
Furthermore, average values of the spectral index at proton scales, based on statistical studies employing
a large data set, would depend on the relative amount of fast and slow wind present in the data set itself.
Further analysis are needed for matching results of numerical simulations and  observational
data, mostly by making more specific the conditions in which the problem is addressed. It should in 
particular be stressed that, according to both the FLR-Landau fluid simulations and the phenomenological model, 
the turbulence state cannot be characterized by the sole amplitude of the turbulence fluctuations, but rather by the 
nonlinearity parameter which is however difficult to estimate from solar wind observations.

\section{Magnetic field line meandering}

\begin{figure*}
\begin{center}
\includegraphics[width=0.325\textwidth]{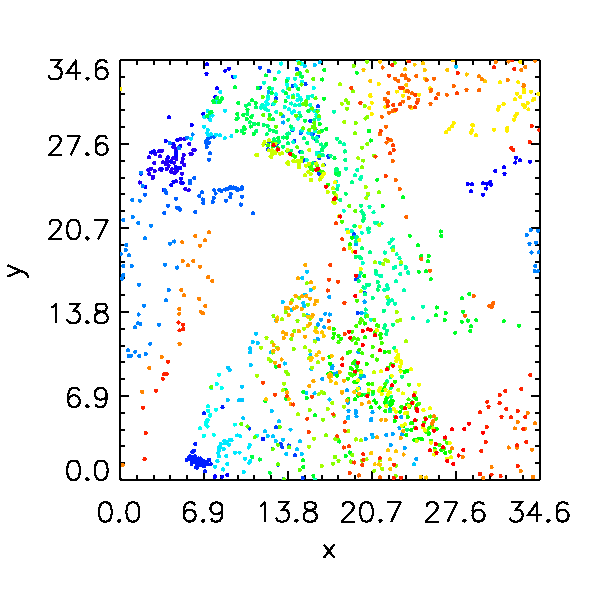}
\includegraphics[width=0.325\textwidth]{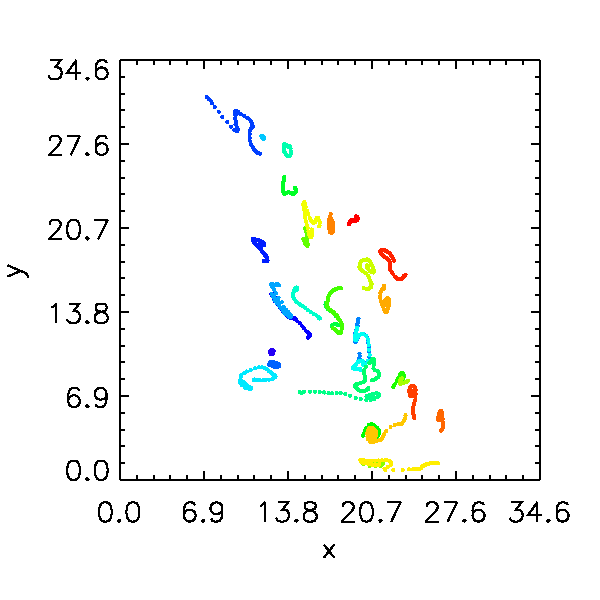}
\includegraphics[width=0.325\textwidth]{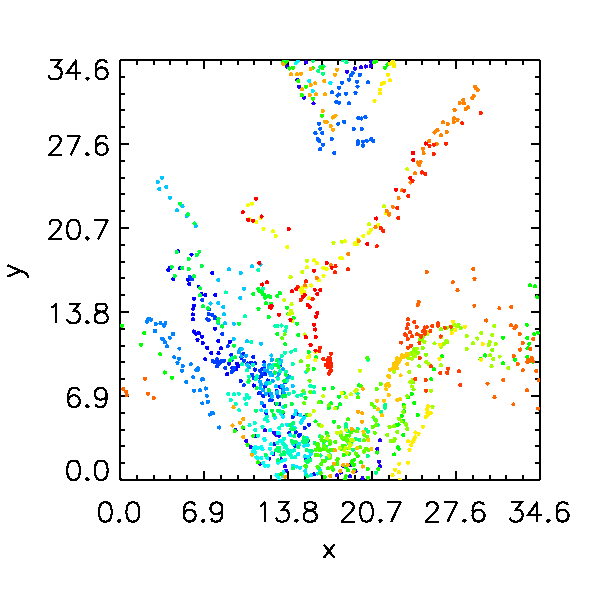}
\end{center}
\caption{Poincar\'e map for runs $R_{0.2}^{80}$ (left), $R_{0.08}^{80}$ (middle)
and $R_{0.08}^{86}$ in the transverse plane $z=L_z/2$, of 32  magnetic 
field lines originating from equidistributed points in a square 
of the plane $z=0$, whose  linear dimensions are about $L_\perp/3$,
at a time when turbulence has reached a stationary regime. Various colors correspond
to the different field lines.}
\label{Pmap}
\end{figure*}

A main observation of Section \ref{spectr} is that the strength of the turbulence and 
the spectral exponent at the sub-ion scales are governed by the nonlinearity parameter
that itself results from the driving conditions, characterized by the parameter $A$ rather than by the 
amplitude of the turbulence fluctuations only. It is thus of interest to also analyze the influence of the 
parameter $A$ in physical space. For this purpose, we explore the 
magnetic field line properties at a given time when turbulence has reached a stationary state. 
Taking advantage of the periodicity of the computational domain, these (non periodic) lines are computed beyond the 
computational domain. Turbulence statistical homogeneity then allows one to 
explore various regions when integration is carried out on many spatial periods. It is then 
convenient to consider the Poincar\'e maps defined by the intersection points of the 
considered magnetic lines with a transverse plane, here chosen to be located at $z= L_\|/2$. These
points can be viewed as the projections of the intersection points of the
field lines with parallel transverse planes separated by a distance $L_\|$.
The erratic character of the intersection points associated with a given line and the mixing of the 
different lines can then be viewed as reflecting the strength of the turbulence.

We have considered Poincar\'e maps for magnetic field lines corresponding to  
runs $R_{0.2}^{80}$, $R_{0.08}^{80}$ and $R_{0.08}^{86}$. The first two runs differ by the amplitude of the 
turbulent fluctuations, while the third one is characterized by the same level of fluctuations as the 
second one but has the same parameter $A$, and essentially the same nonlinearity parameter $\chi$,  as the first one. 
Figure \ref{Pmap} displays the 50 intersections for the 32 considered magnetic lines
(each of them characterized by a different color), in the three 
simulations. Interestingly, although they correspond to a different level of fluctuations, 
runs  $R_{0.2}^{80}$ and $R_{0.08}^{86}$ (that display the same $\chi$) both display an erratic distribution 
of the intersection points of magnetic field lines in the transverse plane and a significant mixing of these
lines, a situation that appears to be associated to a regime of strong turbulence. 
On the other hand,  for run $R_{0.08}^{80}$ that corresponds to a regime of weak turbulence, the 
successive intersection points of individual field lines form distinct clusters,
a configuration which is not significantly affected when increasing the extension of the considered field
lines and thus the number of intersections. 
These observations indicate that the degree of  meandering of the magnetic field lines is not associated
with the amplitude of the turbulence fluctuations but rather to the value of the nonlinearity parameter, which
governs the weak or strong character of the turbulence. This remark could be of interest in the context
of propagation of solar energetic particles (SEP) for which the sole Fokker-Planck equation cannot explain 
the fast longitudinal diffusion \citep{Laitinen15}.

\section{A phenomenological model} \label{pheno-model}

The non universality of the transverse energy spectrum of the magnetic fluctuations 
at the sub-ion scales 
and its sensitivity to the nonlinearity parameter, observed numerically with the FLR-Landau fluid,
is also predicted by the phenomenological model presented in \citet{PS15}, aimed at determining
the stationary magnetic energy spectrum at scales smaller than the transition range if it exists.
We here briefly review this model in the simpler situation where the nonlinear interactions 
are mostly local, which is the case for $\beta=1$ because 
the energy spectrum is not very steep. Ion and electron mean temperatures are assumed equal and
isotropic.

In addition to the nonlinear frequency $\omega_{NL} = \Lambda [k_\perp^5\rho_i^2 E_k]^{1/2}$ (here $\Lambda$
is a numerical constant of order unity, and $E_k$ holds for $E(k_\perp$)), and 
the frequency $\omega_W = {\overline \omega} k_\| v_A$ (where ${\overline \omega}$ scales like $k_\perp\rho_i$)
of KAWs propagating along the (distorted) magnetic field lines, additional characteristic time scales
have to be considered. Landau resonance induces a damping rate, given by $\gamma = {\overline \gamma} k_\| v_A$ where 
${\overline \gamma}$ scales approximately like $k_\perp^2\rho_i^2$, at least for $\beta$ of order unity,
but also a homogenization frequency $\omega_H = \mu k_\| v_{th}$ (where $\mu$ is a proportionality constant),
which in the case of ions is comparable to the 
other inverse characteristic time scales. Due to the mass ratio, the corresponding
frequency is much higher in the case of the 
electrons, making the homogenization of the electron  along the magnetic field lines too fast for having
a significant dynamical effect. This point will be addressed in the next section. The problem then arises of evaluating
the characteristic transfer time, or its inverse $\omega_{tr}$. For this purpose, we proceed in the spirit of the 
classical two-point closures, such as the EDQNM model, classically used in  hydrodynamic turbulence 
\citep{Orszag70,Orszag76,SLF75,Lesieur08} and also in the context of homogeneous MHD where the influence of
large-scale Alfv\'en wave on the inertial dynamics is to be retained \citep{PFL76}. Details are given in the 
Appendix. One is led to write
\begin{equation}
\omega_{tr} = \frac{\omega_{NL}^2}{\omega_W + \omega_H} = 
\frac{\Lambda^2 {\overline \alpha}^2k_\perp^3 E_k}{{\overline \omega}v_A k_\| + \mu v_{th} k_\|} \label{omega-tr}
\end{equation}
where ${\overline \alpha}= {\overline \omega} \sim k_\perp\rho_i$.

One can then define a turbulent energy flux by 
\begin{equation}
\epsilon = C\omega_{tr} k_\perp E_k, \label{epsilon}
\end{equation}
where $C$ is a negative power of the Kolmogorov constant. Substituting
Eq. (\ref{omega-tr}) into Eq. (\ref{epsilon}), one gets
\begin{equation}
\epsilon = \frac{C\Lambda^2 {\overline \alpha}^2 k_\perp^4 E_k^2}{({\overline \omega} v_A + \mu v_{th}) k_\|}.\label{flux}
\end{equation}
At scales for which $k_\perp\rho_i \gg 1$, determining $k_\|$ from the critical balance argument
$\omega_{NL} \propto \omega_W$, it follows that, to leading order (taking, up to a simple rescaling, the 
corresponding proportionality constant equal to $\Lambda$ which then identifies with the 
nonlinearity parameter), 
\begin{equation}
E_k \sim  \Lambda^{-4/3} C^{-2/3} \epsilon^{2/3} k_\perp^{-7/3}. \label{spct}
\end{equation}
Here, due to Landau damping, $\epsilon$ is a function of $k_\perp$ and decays along the cascade. 

Indeed, retaining linear Landau damping leads to the phenomenological equation  for KAW's energy spectrum
\citep{Howes08,Howes11}
\begin{equation}
\partial_t E_k + {\mathcal T}_k = -2\gamma E_k + S_k,
\end{equation}
where $S_k$ is the driving term acting at large scales and ${\mathcal T}_k$ the transfer term 
related to the energy flux  $\epsilon$
by ${\mathcal T}_k = \partial \epsilon/\partial {k_\perp}$. Due to Landau damping, 
energy is not transferred conservatively along the cascade, making  $\epsilon$ 
scale-dependent. For  a steady state and outside the injection range, one has
\begin{equation}
{d\epsilon}/{dk_\perp} = - 2 \gamma E_k \equiv  - 2 {\overline \gamma} v_A k_\| E_k, \label{balance}
\end{equation}
where, using Eq. (\ref{flux}) (after multiplying the numerator and denominator of the RHS by $k_\|$),  
\begin{equation}
k_\| E_k = \frac{C^{-1}\Lambda^{-2}}{k_\perp {\overline \alpha}^2 v_A^2} ({\overline \omega} v_A + \mu v_{th}) 
\left ( \frac{v_A^2k_\|^2}{k_\perp^3E_k}\right) \epsilon.
\end{equation}
In a critically balanced regime, where  $k_\|v_A= (k_\perp^3 E_k)^{1/2}$, Eq.  (\ref{balance}) is then solved as 
\begin{equation}
\epsilon = \epsilon_0 \exp \left[-2C^{-1}\Lambda^{-2} \int_{k_0}^{k_{\perp}} 
\frac{{\overline \gamma}}{\xi {\overline \alpha}^2}
({\overline \omega} + \mu \beta^{1/2})d\xi\right]. \label{epsilonk}
\end{equation}
This (scale-dependent) energy flux is to be substituted in Eq. (\ref{spct}), using 
${\overline \alpha} = {\overline \omega}$ and 
${\overline \gamma}/{\overline \omega}^2 \approx \delta(\beta)$,
which can be approximated by  $0.78 \rho_e/\rho_i$
when using Eqs. (62) and (63) of \citet{Howes06} with $\beta =1$. Furthermore, 
in this range, ${\overline \omega} \approx a(\beta) k_\perp$ with  $a(\beta) = (1+\beta)^{-\frac{1}{2}}$.
This leads to $\epsilon_k \sim \epsilon_0 k_\perp^{-\zeta} \exp [-2 a(\beta) C^{-2}\Lambda^{-2}\delta(\beta) k_\perp]$, 
with $\zeta= 2\delta(\beta) C^{-1}\mu\Lambda^{-2} \beta^{1/2}$.
This results in  a steepening of the algebraic prefactor of the magnetic spectrum 
which is now given by
\begin{equation}
E_k \sim k_\perp^{-(7/3 + 2\zeta/3)} \exp 
\left [-\frac{4}{3} a(\beta)\delta(\beta) C^{-1}\Lambda^{-2} (k_\perp\rho_i)\right].
\end{equation}
We note in particular that increasing the nonlinearity parameter $\Lambda$ makes the power law spectrum shallower.

\section{Landau damping and particle streaming}

\begin{figure*}
\centerline{
\includegraphics[width=0.29\textwidth]{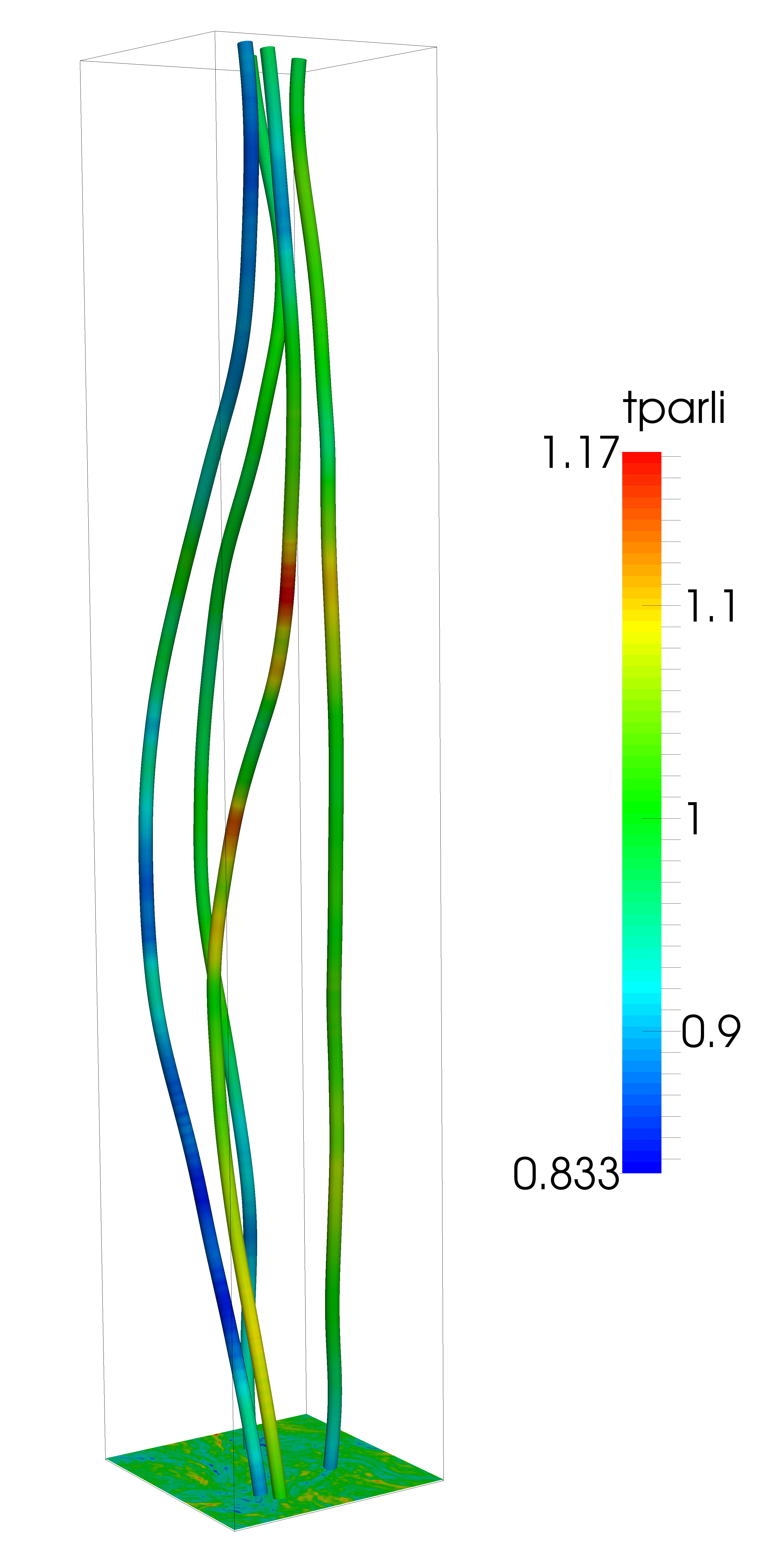}
\includegraphics[width=0.29\textwidth]{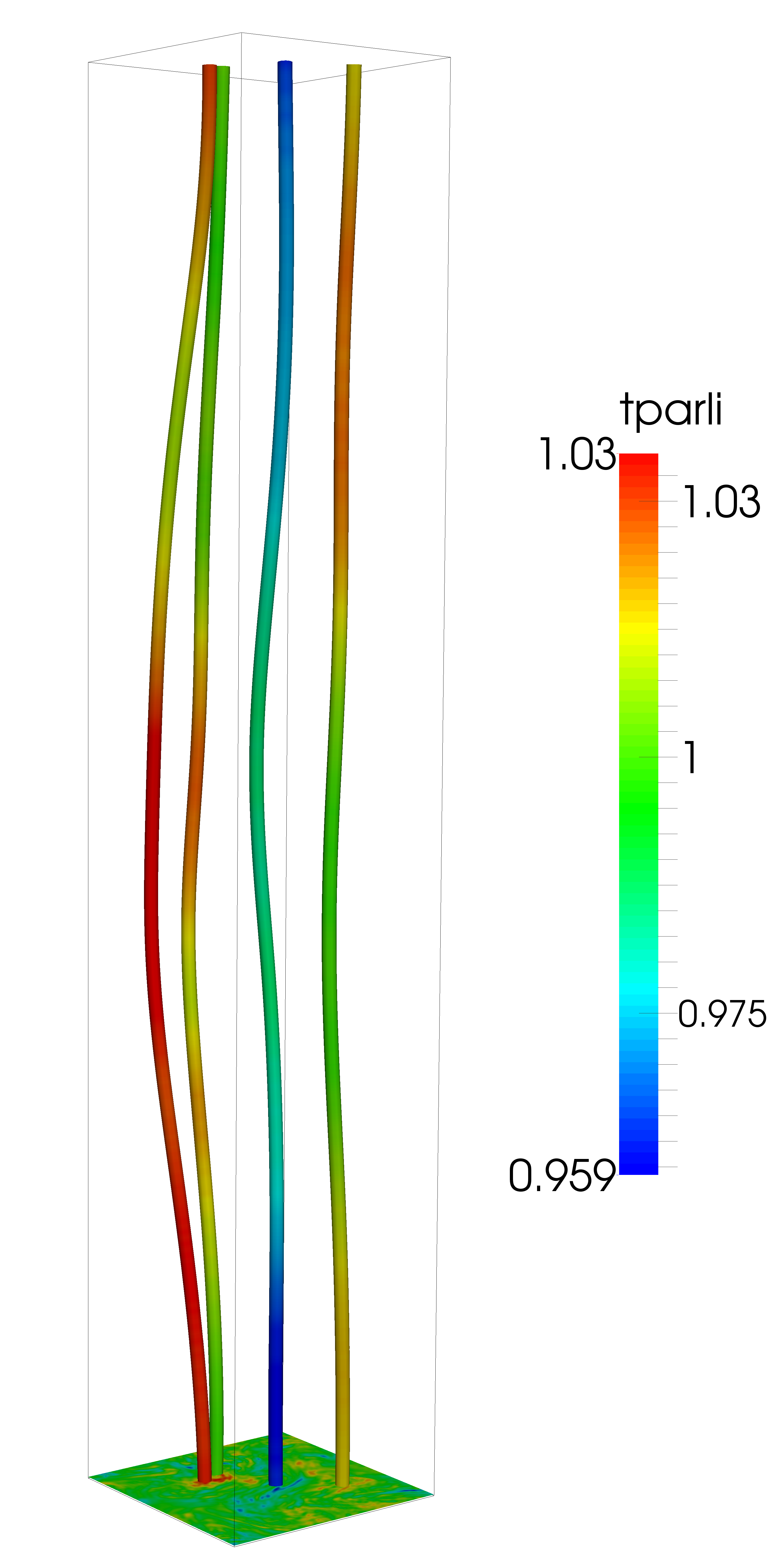}
\includegraphics[width=0.29\textwidth]{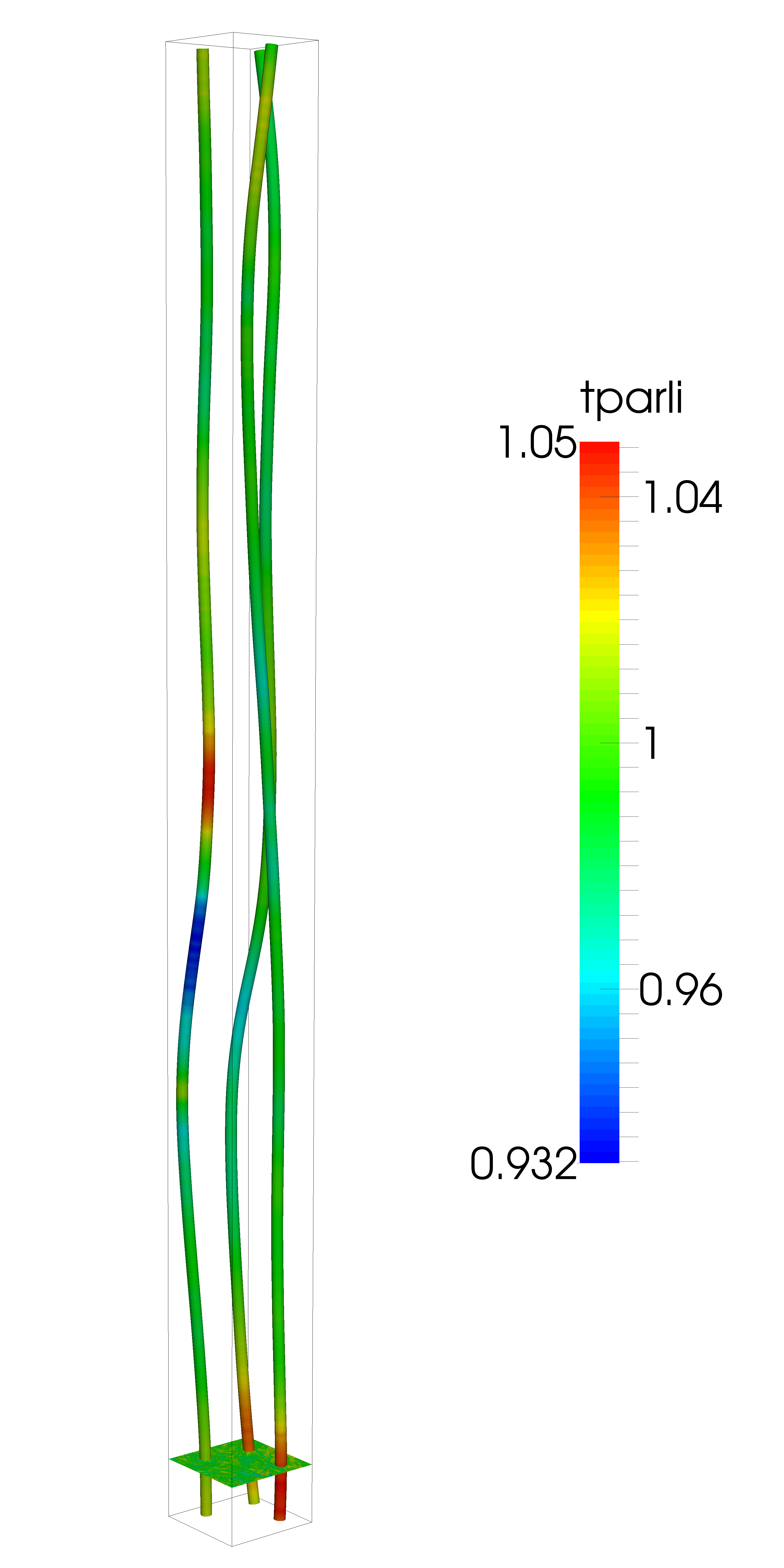}
}
\vskip0.05cm
\centerline{
\includegraphics[width=0.29\textwidth]{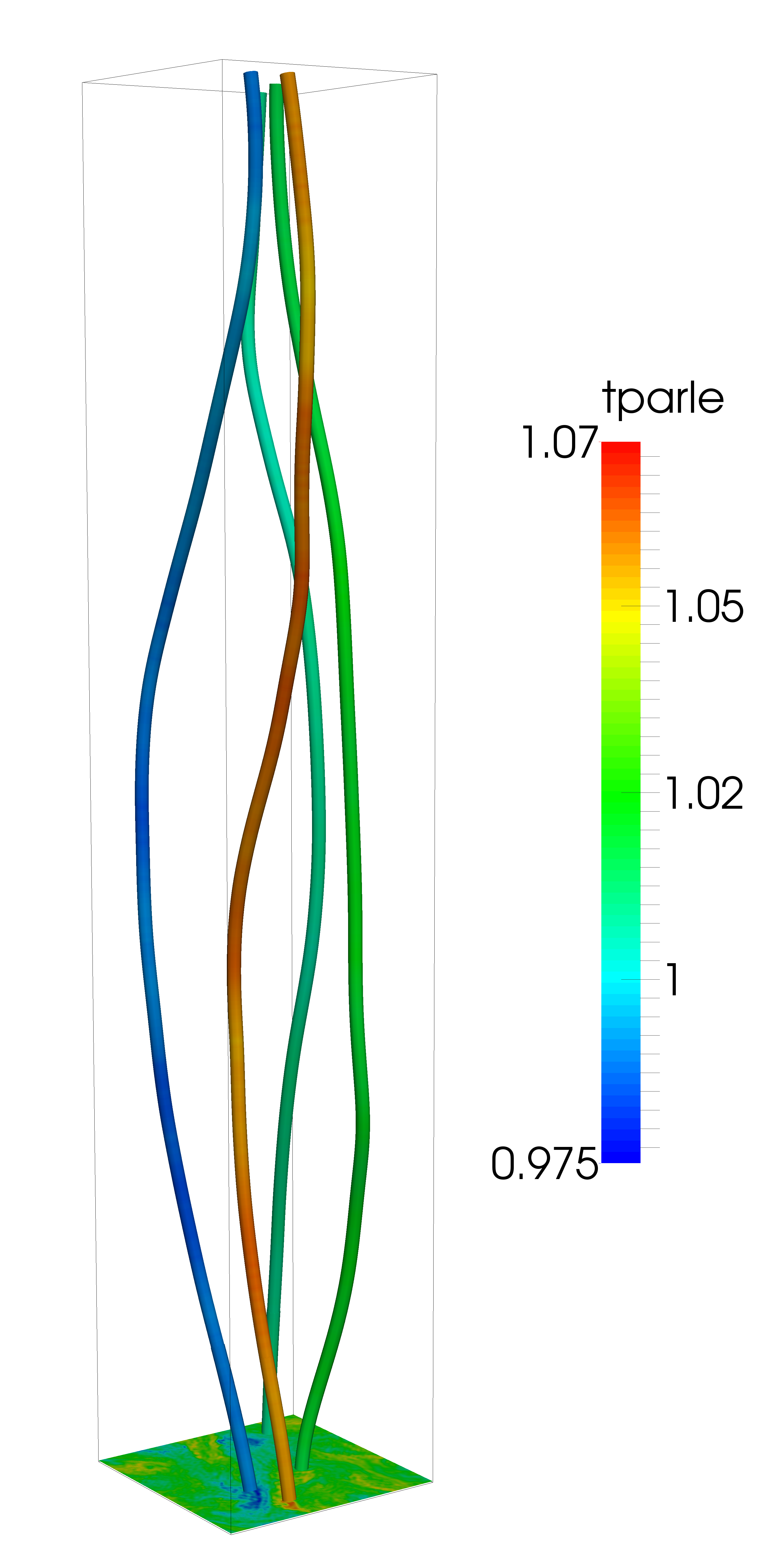}
\includegraphics[width=0.29\textwidth]{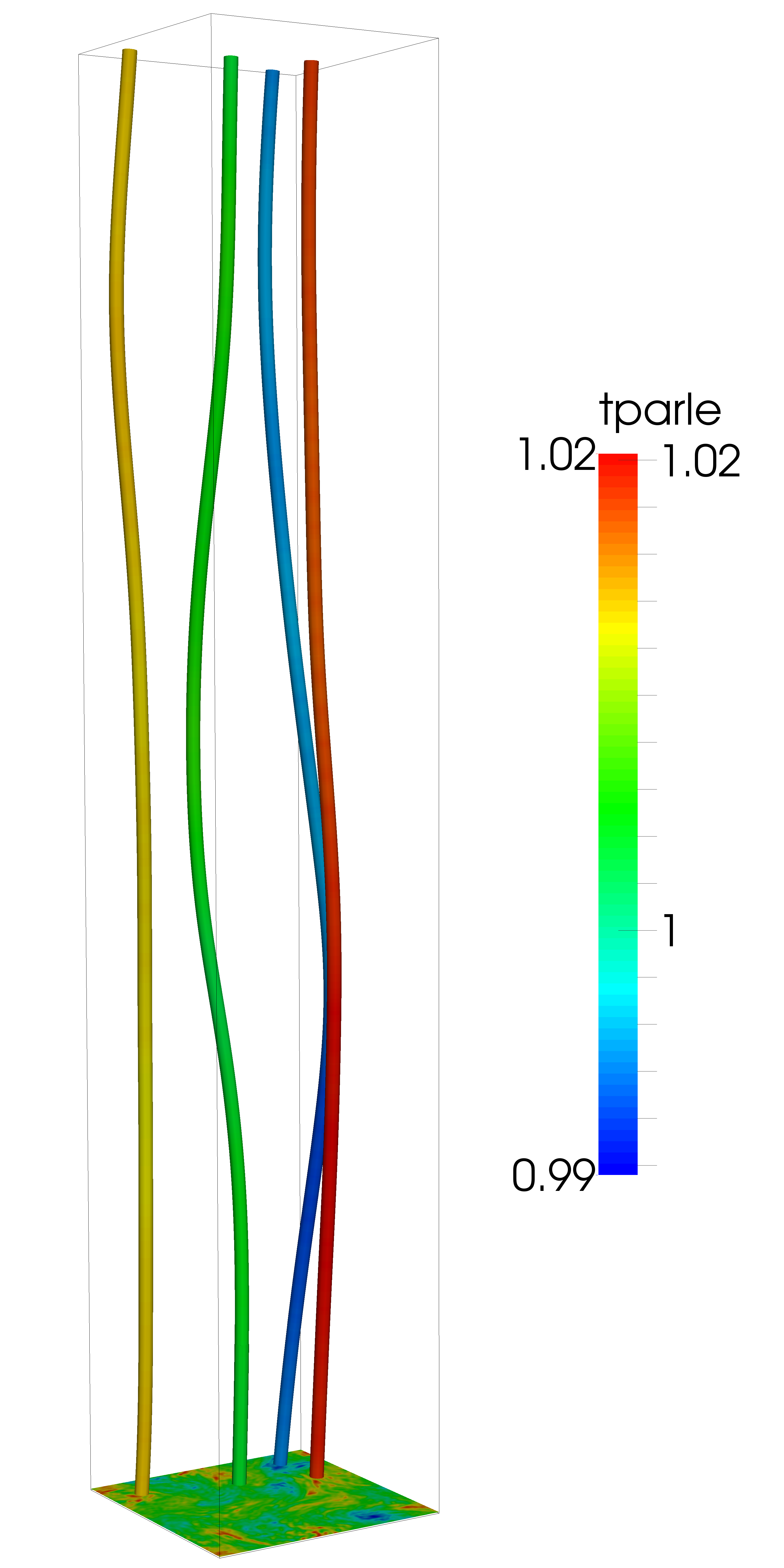}
\includegraphics[width=0.29\textwidth]{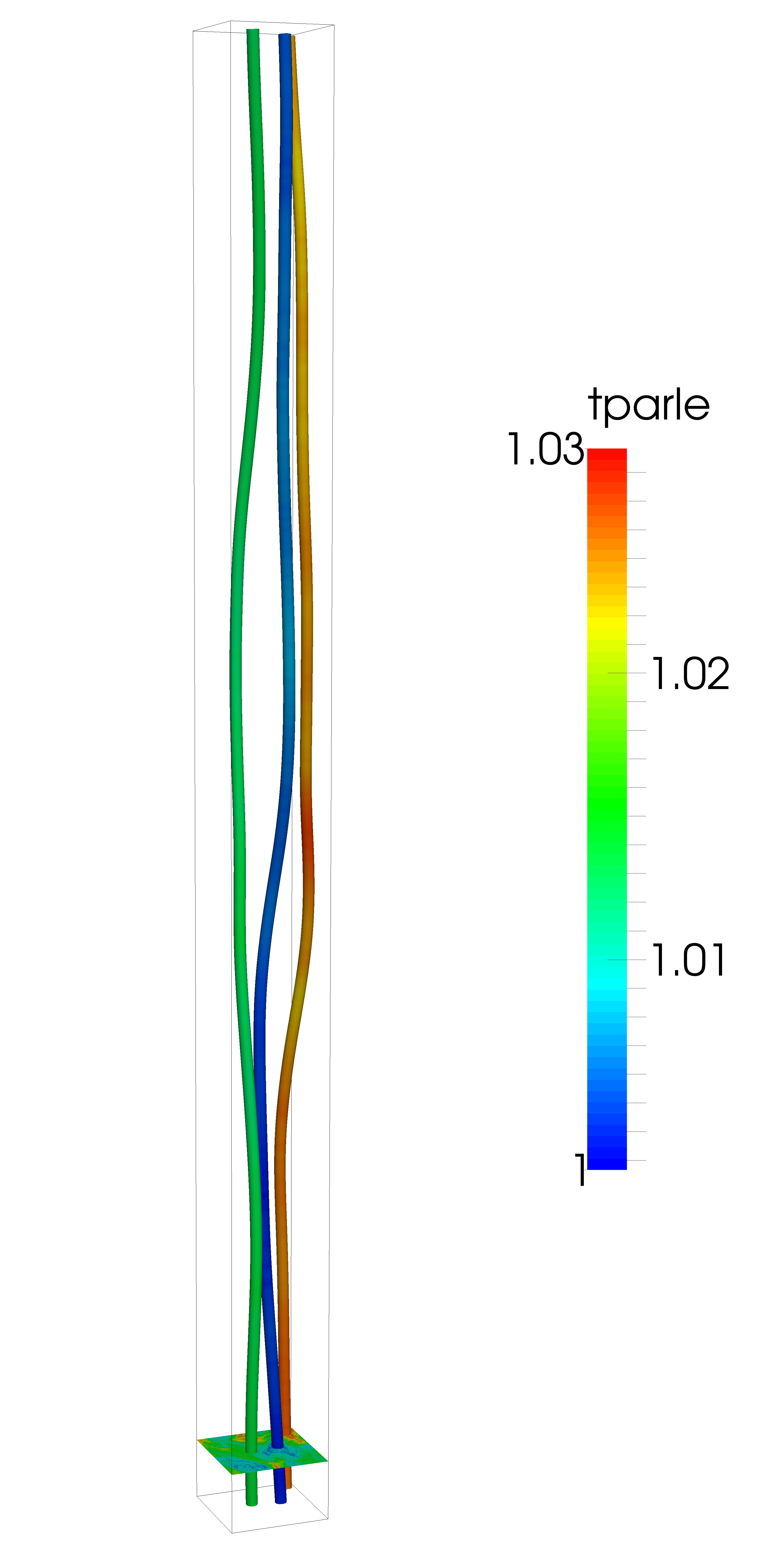}
}
\caption{Magnetic field lines colored in function of the parallel temperature of the ions (top)
and of the electrons (bottom) for the runs $R_{0.2}^{80}$ (left), $R_{0.08}^{80}$ (middle) 
and $R_{0.08}^{86}$ (right), when turbulence has reached a stationary state. 
Different field lines are displayed for the ions and the electrons in order to make more 
visible the variations of the electron temperature from one line to the other. 
}
\label{Filines}
\end{figure*}

\begin{table*}
\begin{center}
\def~{\hphantom{0}}
\begin{tabular}[b]{|c||c|c|c|c|c|c|}
\hline
     &     \multicolumn{2}{|c|}{}           &        \multicolumn{2}{|c|}{}    & \multicolumn{2}{|c|}{}          \\
 Run & \multicolumn{2}{|c|}{$R_{0.2}^{80}$} &  \multicolumn{2}{|c|}{$R_{0.08}^{80}$} &   \multicolumn{2}{|c|}{$R_{0.08}^{86}$} \\ 
     &      \multicolumn{2}{|c|}{}       &     \multicolumn{2}{|c|}{} &  \multicolumn{2}{|c|}{}                \\
\hline
$T_{\|i}$ & $\sigma_s^2 =1.5 \times 10^{-3}$ & $L_s =13.0$ & $\sigma_s^2 = 5 \times 10^{-5}$ & $L_s =22.4$ & $\sigma_s^2 = 1.4 \times 10^{-4}$ & $L_s =34.2$ \\
     & $\sigma_z^2 =2.2 \times 10^{-3}$ & $L_z =10.1$ & $\sigma_z^2 = 8 \times 10^{-5}$ & $L_z =15.9$ &  $\sigma_z^2 = 1.5 \times 10^{-4}$ & $L_z =12.3$       \\
\hline
$T_{\|e}$ & $\sigma_s^2 =2.2 \times 10^{-5}$ & $L_s =23.6$ & $\sigma_s^2 =  1.1 \times 10^{-6}$ & $L_s =22.7$ & $\sigma_s^2 = 1.3 \times 10^{-6}$ & $L_s =53.5$\\
          & $\sigma_z^2 = 1.8 \times 10^{-4}$ & $L_z =16.5$ & $\sigma_z^2 = 9.0 \times 10^{-6}$ & $L_z =16.0$ & $\sigma_z^2 = 1.3 \times 10^{-5}$  & $L_z = 44.6$ \\
\hline
$T_{\perp i}$ & $\sigma_s^2 =2.2 \times 10^{-3}$ & $L_s =17.2$ & $\sigma_s^2 = 2.2 \times 10^{-4}$ & $L_s =26.5$ & $\sigma_s^2 = 2.7 \times 10^{-4}$ & $L_s =42.0$ \\
            & $\sigma_z^2 =2.3 \times 10^{-3}$ & $L_z =13.9$ & $\sigma_z^2 = 2.3 \times 10^{-4}$ & $L_z =19.8$ &  $\sigma_z^2 = 2.6 \times 10^{-4}$ & $L_z =25.0$       \\
\hline
$T_{\perp e}$ & $\sigma_s^2 =2.2 \times 10^{-5}$ & $L_s =24.3$ & $\sigma_s^2 =  1.7 \times 10^{-6}$ & $L_s =28.0$ & $\sigma_s^2 = 3.2 \times 10^{-5}$ & $L_s =67.7$\\
          & $\sigma_z^2 = 5.9 \times 10^{-5}$ & $L_z =18.5$ & $\sigma_z^2 = 6.4 \times 10^{-6}$ & $L_z =16.7$ & $\sigma_z^2 = 4.4 \times 10^{-5}$  & $L_z = 70.0$ \\
\hline
$|B|^2$ &$\sigma_s^2 =7.8  \times 10^{-3}$ & $L_s = 16.2$ & $\sigma_s^2 =  8.9 \times 10^{-4}$ & $L_s =25.1$ & $\sigma_s^2 = 1.1 \times 10^{-3}$ & $L_s =46.3$\\
          & $\sigma_z^2 =9.4 \times 10^{-3}$ & $L_z =9.2$ & $\sigma_z^2 = 8.2 \times 10^{-4}$ & $L_z = 17.2 $ & $\sigma_z^2 = 1.1 \times 10^{-3}$ & $L_z = 24.5 $ \\
\hline
$|B_\perp|^2$ &$\sigma_s^2 = 1.8 \times 10^{-3}$ & $L_s = 14.3$ & $\sigma_s^2 = 2.7 \times 10^{-5}$ & $L_s =15.0$ & $\sigma_s^2 = 5.9 \times 10^{-5}$ & $L_s = 34.0$\\
          & $\sigma_z^2 = 2.8  \times 10^{-3} $ & $L_z = 14.1$ & $\sigma_z^2 =2.4 \times 10^{-5} $ & $L_z = 15.3$ &  $\sigma_z^2 = 5.0 \times 10^{-5}$ & $L_z=36.9$\\
\hline
\end{tabular}
\caption{Variance of the ion and electron parallel temperatures  along magnetic field lines ($\sigma_s^2$) and along  parallels to the $z$-axis
($\sigma_z^2$), together with the corresponding correlation lengths $L_s$ and $L_z$ for runs
$R_{0.2}^{80}$,  $R_{0.08}^{80}$ and  $R_{0.08}^{86}$.}
\end{center}
\end{table*}

A main feature of the phenomenological model discussed in Section \ref{pheno-model} concerns the role
played by the homogenization of quantities like temperatures along the magnetic field lines, through an effect of
particle streaming. The aim of this section is to get some insight into this process.
One may expect that it is more efficient for the electrons than for the ions, because of the mass ratio.
The question also arises of the influence of the amplitude of the turbulence fluctuations.
This issue is qualitatively exemplified in Fig. \ref{Filines} which
displays a few magnetic field lines colored according to the parallel temperature of the ions (top)
and of the electrons (bottom) for the runs $R_{0.2}^{80}$ (left), $R_{0.08}^{80}$ (middle) and $R_{0.08}^{86}$
(right). We indeed observe that the parallel electron temperature  is significantly more uniform than 
the parallel ion temperature and that the homogenization process is more efficient  
for run $R_{0.2}^{80}$ than for run $R_{0.08}^{80}$ corresponding to a smaller nonlinearity parameter. 
The significant inhomogenity of the ion temperature along the magnetic field
lines of run $R_{0.08}^{86}$ demonstrates that $\chi$ rather than the level 
of the turbulence fluctuations prescribes the dynamics.

These observations can be made more quantitative by computing
the  variances $\sigma^2_S$ and  $\sigma^2_z$ and the correlation lengths $L_s$ and $L_z$
of the ion and electron parallel and perpendicular temperatures, averaged 
over the 32 field lines mentioned in Section \ref{magneticfieldlines}
(subscript $s$) and over the corresponding lines parallel to the z-axis (subscript $z$).
Table 2 collects data concerning run $R_{0.2}^{80}$,  $R_{0.08}^{80}$ and  $R_{0.08}^{80}$.
For comparison, variances and correlation lengths are also presented for the magnetic field amplitude $|B|$ 
and the transverse component $|B_\perp|^2$.

Table 2 clearly shows that temperature fluctuations as measured by their respective  variances
are more important for the ions than for the electrons, and decrease with the amplitude of the
magnetic fluctuations, even for comparable values of the nonlinearity parameter $\chi$. 
The temperature variances along field lines are usually smaller than along the z-direction, 
except for the perpendicular ion temperature which is less sensitive to Landau damping.

For each run and each particle species, the temperature correlation lengths $L_s$ and $L_z$  
are  defined as the integral up to the first zero of the normalized correlation function
$\langle T_\|({\mathbf x'} + {\mathbf x)} T_\|({\mathbf x'})\rangle/\langle T_\|({\mathbf x})^2\rangle$ 
computed along each of the 32 selected lines and averaged over them. 
When considering the ideal case of a sinusoidal fluctuation of wavenumber $k= 1$, this definition
leads to a correlation length equal to one. A correlation $L_s$ along the magnetic field
lines will thus correspond to a parallel wavenumber $K_\|= 1/L_s$. Similar definitions are
given in the case of the fields $|B|^2$ and $|B_\perp|^2$. It is of interest to compare the
quantity $K_\|$ evaluated for $|B|^2$,  with the scale-dependent parallel wavenumber 
$k_\|$ previously discussed. From Table 2, the values $K_\|$ corresponding to runs $R_{0.2}^{80}$,  $R_{0.08}^{80}$
and $R_{0.08}^{86}$ are $0.06$, $0.04$ and $0.02$ respectively, which from Fig. \ref{kparal-fig}
roughly corresponds to values
of $k_\|$ between $k_\perp=0.4$ and $0.6$, a range where the transverse magnetic spectrum
starts displaying a power law (Fig. \ref{chi-spec}, bottom).
The correlation length of $|B|^2$ in the z-direction is significantly smaller than along the field lines,
in contrast with the variances which are of the same order of magnitude. This variation is due to the $B_z$
component, as for $|B_\perp|^2$ both variances and correlations lengths are similar 
in both directions. The correlation lengths of  $|B_\perp|^2$ are indeed prescribed by the 
KAW wavelength, and are thus essentially independent of the amplitude of the turbulent fluctuations.
When comparing the correlations length for runs  $R_{0.2}^{80}$ and $R_{0.08}^{86}$, one recovers the
$2.5$ geometric scaling factor between the longitudinal extensions of the computational domains,
which is also the ratio of the respective parallel wavelengths of the driven waves.

Considering electron temperatures, the correlation lengths along the field lines are 
roughly the same for the three runs, up to the $2.5$ scaling factor for $R_{0.08}^{86}$. 
They are significantly larger than in the z-direction, indicating that Landau damping leads to
homogenization along the field lines, but not to isothermal electrons in the full domain.
Concerning the ion temperatures, the two runs with the same nonlinearity parameter
have similar correlation lengths up to the geometric factor (larger along the field lines than 
in the z-direction). In contrast, for the run with a smaller value of $\chi$, the correlation 
lengths are larger.

To summarize, these results show that Landau damping, as implemented in the FLR-Landau fluid code, 
induces a homogenization  of 
temperature fluctuations along the field lines. They also comfort the assumption of the phenomenological model 
that the electron homogenization time scale can be ignored, as this process appears to be essentially independent of the 
turbulence characteristics. The ion temperature correlation length normalized by 
the parallel wavelength of the driven waves is sensitive to the nonlinearity parameter, suggesting that 
the nonlinear dynamics is indeed coupled to the ion homogenization process. The influence of 
this effect on the energy transfer is permitted by the fact that the frequencies $\omega_H \sim v_{th} k_\|$ and 
$\omega_{NL} = k_\perp v_e \sim [k_\perp^5 E_k]^{1/2}$
are of the same order of magnitude near the ion gyroscale. Using the fact that  $v_{th} \sim v_A \sim 1$,
we indeed check that, for run $R_{0.2}^{80}$, the ratio   $\omega_{NL}/\omega_H$ is $0.34$
at $k_\perp=1.1$, $0.55$ for $k_\perp=2$ and $0.9$ for $k_\perp=2.9$, while for run $R_{0.08}^{80}$ this ratio 
takes the values $0.19$, $0.25$ and $0.34$ for the corresponding values of $k_\perp$. This indicates that
Landau damping is more efficient in the latter run which displays the steeper spectrum. For comparison,
in the case of run $R_{0.08}^{86}$, this ratio takes the values $0.33$, $0.56$ and $0.84$, rather
similar to those of run  $R_{0.2}^{80}$ that is characterized by the same nonlinearity parameter.

\section{Conclusion}

The three-dimensional FLR-Landau fluid simulations of sub-ion turbulence in the solar wind 
analyzed in this paper show a non-universal power-law spectrum for the transverse magnetic energy,
with an exponent depending  on the nonlinearity parameter, a quantity 
which is varied by changing
the propagation angle and the relative amplitude of the driven KAWs.
The nonlinearity parameter, rather than the amplitude of the magnetic fluctuations, 
prescribes the turbulence strength, governing for example the meandering 
of the magnetic field lines. The simulation results are supported by 
a phenomenological model where the departure from the usually predicted 
$k_\perp^{-7/3}$ spectrum originates from the sensitivity of the transfer time to the 
particle streaming along the magnetic field lines induced by Landau damping.
Interestingly, similar simulations where Landau resonance is eliminated by 
prescribing zero heat fluxes (bi-adiabatic approximation) leads to a $-7/3$
exponent, whatever the value of the nonlinearity parameter. Details will be given elsewhere. 

The FLR-Landau fluid simulations also clearly point out the importance of 
the dynamics along the field lines and in particular the 
anisotropic homogenization process, which contrasts with the often used 
assumption of isothermal electrons. 

The present simulations, together with the phenomenological modeling, provide an 
interpretation of the significant dispersion of the sub-ion spectral exponent
measured in the solar wind. An interesting development would be to evaluate the
nonlinearity parameter from satellite data. This turns out to be a difficult 
but important issue, especially because correlation 
of the spectral exponent with 
the sole amplitude of the turbulence fluctuations in various regions of the solar wind
leads to different conclusions.

Although the present simulations seem sufficient to characterize the energy spectrum
of the transverse magnetic fluctuations at the sub-ion scales, higher resolutions are
in project to capture part of the MHD range and also to increase the extension of the
simulated sub-ion spectral domain. The latter development is important to reach 
scales where the electron-MHD approximation starts to apply.

\acknowledgments 
The research leading to these results has received funding from the 
European Commission's Seventh Framework Programme (FP7/2007-2013) under
the grant agreement SHOCK (project number 284515). 
The numerical simulations were performed using 
high-performance computing resources from GENCI-IDRIS (Grant i2013047042) and 
computing facilities provided by the ``Mesocentre SIGAMM'' hosted by 
Observatoire de la C\^ote d'Azur.

\appendix
\section{Estimate of the characteristic energy transfer time}

Writing the Navier-Stokes equation, or any quadratically nonlinear initial value problem,
(neglecting for the sake of simplicity driving and dissipation)
in the symbolic form $\partial_t u = uu$ and taking the statistical moments, we are 
led to write (assuming spatial homogeneity)
\begin{eqnarray}
&&\partial_t \langle uu \rangle = \langle  uuu \rangle \label{uu}\\
&&\partial_t \langle uuu \rangle = \langle  uuuu \rangle \equiv 
  \langle uu \rangle  \langle uu \rangle  + \langle  uuuu \rangle_c \label{uuu}
\end{eqnarray}
where we separate the Gaussian contribution symbolically written  $\langle uu \rangle  \langle uu \rangle$,
from the fourth-rank cumulant $\langle  uuuu \rangle_c$. Neglecting this latter contribution would  correspond 
to the quasi-normal approximation, where the unlimited growth of the third-rank correlations leads to 
unphysical effects such as negative energy spectra \citep{Ogura63, OBrian62, Orszag76}. 
The fourth-order cumulant, providing a  coupling with the higher moments
of the fluid hierarchy, actually ensures saturation of these correlations. By an argument similar to that 
used to introduce the concept of  turbulent viscosity, we write $\langle  uuuu \rangle_c = - \eta \langle uuu\rangle$.
After a transient, we thus get a quasi-equilibrium state where 
$ \langle  uuu \rangle = \eta^{-1}  \langle uu \rangle  \langle uu \rangle$, which is substituted in Eq. (\ref{uu})
that becomes 
\begin{equation}
\partial_t \langle uu \rangle = \eta^{-1}  \langle  uu \rangle \langle  uu \rangle. 
\end{equation}
This equation is in fact an integrodifferential equations involving wavevector triads. Here, we will drastically
simplify this model by assuming strict locality of the interactions. When turning to the energy spectrum, we
are then led to write
\begin{equation}
\partial_t E_k = \eta^{-1} \omega_{NL}^{2} E_k,
\end{equation}
on a dimensional basis, noting that only energy spectrum and wavenumber are entering the equation.
We can then define an energy  transfer time $\tau_{tr}= \omega_{tr}^{-1}$ with 
$\omega_{tr}= \eta^{-1} \omega_{NL}^2$,
where the relaxation rate $\eta$ of triple correlations is estimated as the sum of the frequency of the
contributing processes.  In purely hydrodynamic turbulence
$\eta=\omega_{NL}$, so $\omega_{tr} = \omega_{NL}$. In standard (weak) wave turbulence where $\omega_W \gg \omega_{NL}$,
we  have $\omega_{tr} = \omega_{NL}^2/\omega_W$. Here, we also have to take into account the contribution of the
homogenization process along the field lines of frequency $\omega_H$. Note that, except in pure hydrodynamic
turbulence, it is not necessary to retain $\omega_{NL}$ in the definition of $\eta$ as it either negligible
compared with $\omega_W$ when the turbulence is weak or displaying the same $k_\perp$ scaling in the case of 
critical-balanced turbulence. This leads us to write
\begin{equation}
\omega_{tr} = \frac{\omega_{NL}^2}{\omega_W + \omega_H} = 
\frac{\Lambda^2 {\overline \alpha}^2k_\perp^3 E_k}{{\overline \omega}v_A k_\| + \mu v_{th} k_\|}
\end{equation}
where ${\overline \alpha}= {\overline \omega} \sim k_\perp\rho_i$, 
while $\Lambda$ and $\mu$ are numerical constant.


\end{document}